# Multilayer GZ/YSZ Thermal Barrier Coating from Suspension and Solution Precursor Plasma Spray

K. Leng[*], A. Rincon Romero, T. Hussain[*]

*Coatings and Surface Engineering, Faculty of Engineering, University of Nottingham, NG7 2RD*

*+44 115 951 3795, Tanvir.Hussain@nottingham.ac.uk*

**Abstract**

Gas turbines rely on thermal barrier coating (TBC) to thermally insulate the nickel-based superalloys underneath during operation; however, current TBCs, yttria stabilised zirconia (YSZ), limit the operating temperature and hence efficiency. At an operating temperature above 1200 °C, YSZ is susceptible to failure due to phase instabilities and CMAS (Calcia-Magnesia-Alumina-Silica) attack. Gadolinium zirconates (GZ) could overcome the drawback of YSZ, complementing each other with the multi-layer approach. This study introduces a novel approach utilising axial suspension plasma spray (ASPS) and axial solution precursor plasma spray (ASPPS) to produce a double-layer and a triple-layer TBCs with improved CMAS resistance. The former comprised suspension plasma sprayed GZ and YSZ layers while the latter had an additional dense layer deposited through a solution precursor to minimise the columnar gaps that pre-existed in the SPS GZ layer, thus resisting CMAS infiltration. Both coatings performed similarly in furnace cycling test (FCT) and burner rig testing (BRT). In the CMAS test, triple-layer coating showed better CMAS reactivity, as evidenced by the limited CMAS infiltration observed on the surface.

## 1. Introduction

The implementation of a net-zero economy has encouraged the development of more efficient gas turbine engines. Based on Carnot cycles, the efficiency of a combustion engine is directly related to the turbine entry temperature (TET) [1]. Thus, the higher the TET, the higher the efficiency of a gas turbine engine [2]. With the higher operating temperature, the bare nickel-based superalloy components may have reached the melting temperature, resulting in the risk of creep failure over an extended period. The implementation of thermal barrier coatings (TBCs) on these components helped to improve their lifetime. TBCs are a thermal insulation layer that protects the underlying metallic substrates from the harsh environment by reducing the surface temperature of the components in the range of 100 – 300 ºC [3].

In general, a TBC system is comprised of a substrate, a bond coat and a ceramic topcoat. The bond coat, platinum/nickel aluminide or MCrAlY (M = Co or Ni), is a metallic layer that aims to minimise the thermal strain between the substrate and the topcoat and improving oxidation resistance of the underlying substrate, thereby enhancing the durability of the coating [4]. The ceramic topcoat is mainly zirconia-based ceramics, yttria stabilised zirconia (YSZ), which have been developed and employed since the 1970s [5]; however, YSZ encounters phase transformations beyond 1200 ºC, subsequently inducing a high-level stress into the topcoat [6]. The induced stress will speed up the spallation of the topcoat, and reveal the underlying substrate, resulting in a catastrophic failure of the safety-critical component. In addition to that, the YSZ is prone to CMAS (Calcia-Magnesia-Alumina-Silica) attack at 1200 ºC or higher, leading to early spallation of the topcoat. The drawback of the current TBC (i.e., YSZ) initiated the search for the next-generation TBC materials, thereby shifting the attention towards rare-earth (RE) zirconates, either the lanthanum zirconate (LZ) or the gadolinium zirconate (GZ). The RE zirconates gained attention to overcome the drawbacks of YSZ because of their lower thermal conductivity, the higher phase stability at elevated temperatures and the ability to react with CMAS to form a protective apatite phase; however, RE zirconates have a lower fracture toughness than YSZ, resulting in a poor thermal cycling life for a single-layer system. Hence, a multi-layer approach (with underlying YSZ) has been proposed to overcome this drawback [7–10]. Comparing the LZ and the GZ, the GZ has a higher thermal expansion coefficient and a lower thermal conductivity [11]. On the other hand, the double-layer GZ/YSZ system was reported to perform slightly better than the double-layer LZ/YSZ system in thermal cycling tests, judging from the aspect of the TGO growth and spallation behaviour of the two coatings [12].

Air plasma spray (APS) and electron beam physical vapour deposition (EB-PVD) are the most commonly used techniques to deposit TBCs. The former method is used to deposit large and static components (i.e., nozzle guide vanes, combustor tiles in aero engines) while the latter one is used to deposit the rotating components (i.e., high pressure turbine blades). Owing to the columnar microstructure, the TBC produced by EB-PVD method is one of the most durable TBCs as it offers a good strain-tolerance capability and thermal shock behaviour [13,14]; however, EB-PVD deposition method has a lower deposition rate (i.e., ~ 3.4 – 10 µm/min [15,16]) than other thermal spray methods and it requires expensive vacuum chamber and significant installation costs (> £ 10 million per unit). In addition, EB-PVD TBCs also tend to have a higher thermal conductivity (1.5 W m$^{-1}$ K$^{-1}$) than APS TBCs (0.9 W m$^{-1}$ K$^{-1}$) [17]. As an alternative to the EB-PVD deposition method, the comparable columnar structure associated with a low thermal conductivity (typically < 1.5 W m$^{-1}$ K$^{-1}$) can only be produced by the suspension plasma spray (SPS) deposition method, which makes use of submicron-sized powder feedstocks suspended in a liquid medium. The columnar coating structure produced by the SPS has been studied extensively and proven to be similar to or better than the one produced by the EB-PVD process [17,18]. Mahade et al. reported that the lifetime of the SPS TBCs is comparable to the lifetime of the EB-PVD TBCs [19–25]. On the other hand, Jiang et al. produced a double-layer system through the solution precursor plasma spray (SPPS) method, in which the feedstock is produced by mixing solutes and solvents. The SPPS as-sprayed TBC presented a dense vertically crack (DVC) structure with layered porosities (inter-pass boundaries), showing promising thermal cycling lifetime and performance in CMAS attack [26,27].

To determine the better coating structure in terms of thermal cycling lifetime, Ganvir et al. compared SPS and SPPS TBCs with the conventional APS TBCs [28]. Both SPS and SPPS TBCs were reported to have a comparable lifetime to the conventional APS TBCs, but an improved thermal cycling performance for both SPS and SPPS TBCs can be achieved through optimising the coating structures, respectively. On the other hand, Kumar et al. had compared the SPPS DVC structure with the APS lamellar structure, indicating that the SPPS TBC had a better thermal cycling lifetime than the APS TBC, mainly attributed to the DVC structure in the SPPS TBC to accommodate the strain-tolerance in the coating [29]. From these studies, it can be concluded that an optimised columnar structure is the most favourable coating structure for thermal cycling, followed by the DVC structure [28,29]. Although the columnar or DVC structure is favourable in terms of the thermal cycling lifetime, these structures have reduced protection against CMAS infiltration. Instead, these structures act like a pathway for CMAS infiltration, leading to a catastrophic failure of the coatings [6,30–44]. On the contrary, a columnar gap or crack width in the range of 1 – 2 μm could effectively slow down the CMAS infiltration, thereby improving the performance in CMAS attack [40,45]. It is also worth to note that a narrow columnar or crack will induce a higher capillary pressure, easing in CMAS flow progression; however, the higher contact surface area per unit length in the narrow columnar or crack also induces frictional drag to the flow progression of CMAS

[46]. Hence, it can be said that the performance in stopping CMAS infiltration for either the columnar gap or DVC structure is highly dependent on the number of column or crack density and the width of columnar gaps or crack channels on the surface of the topcoat. If the frictional effect dominates over the capillary pressure in the columnar or crack channel, the CMAS infiltration rate can be significantly reduced, resulting in a lower infiltration depth over time [47]. Hence, the stiffening effect of TBCs due to CMAS infiltration will be reduced, prolonging the lifetime of TBCs.

Concluding from the previous studies, it was clear that open porosity at the surface of the topcoat remains the main cause of CMAS infiltration, resulting in TBC failure. Therefore, it is hypothesised that the CMAS resistance of the TBC could be improved by sealing these open porosities at the top surface of the TBC while still maintaining the strain-tolerance capability of the coating. In this study, a double-layer and a triple-layer coating system were deposited through the SPS and SPPS methods. The double-layer coating was comprised of YSZ and GZ layers, in which both layers were produced with the suspension feedstock. In the triple-layer coating, the coating had an additional thin dense GZ layer (~ 55 µm) that was produced from the solution precursor (SPPS) feedstock [26,48]. The dense SPPS GZ layer aims to seal all the open porosities at the top surface of the topcoat, if not minimise the columnar gaps from the SPS GZ layer. Without these open porosities, the CMAS has a limited pathway to infiltrate the topcoat entirely, protecting the topcoat from CMAS attack. Meanwhile, the preserved columnar gaps in the YSZ and GZ layers could still maintain the strain-tolerant capability of the topcoat.

## 2. Experimental Methods

### 2.1. Substrate and bond coat preparation

Inconel 718 coupons with a nominal composition Ni-19.0Cr-3.0Mo-5.1Nb-0.5Al-0.9Ti-18.5Fe-0.04C (in wt.%) were used as substrates. The substrate had a thickness of 3 mm and a diameter of 12.7 mm. Before the deposition of the topcoat, all substrates were grit blasted (Guyson, Dudley, UK) with fine F100 brown alumina (0.125 – 0.149 mm) particles at 6 bars. The substrates were then cleaned in Industrial Methylated Spirit (IMS) with an ultrasonic bath for approximately 4 minutes. The CoNiCrAlY bond coat (CO-210-24, Praxair, Swindon, UK) was then deposited onto the substrates with High Velocity Oxy-Fuel (HVOF) thermal spray using a commercial Metjet IV (Metallisation, Dudley,

UK) gun [49]. The standard raster scan pattern was achieved with a 6-axis robot (ABB® IRB 2400, Warrington, UK) at a scan speed of 1000 mm/s and a line spacing of 4 mm to factor in the nozzle diameter. A detailed deposition of the bond coat was described in [48,49]. All the bond coat deposition was carried out in the same batch with a thickness value of ~ 110 ± 20 μm.

2.2. Suspension and Solution Precursor Preparation

Two ethanol-based suspensions were supplied by Treibacher Industrie AG (Althofen, Austria). The first suspension was the 8 wt.% yttria stabilised zirconia (AuerCoat YSZ) with a median particle size ($D_{50}$) of 0.45 μm. The second was a gadolinium zirconate (AuerCoat Gd-Zr) with a median particle size ($D_{50}$) of 0.50 μm. Based on the supplier, both suspensions had a solid loading of 25 wt. %. To ensure a well dispersed suspension, both suspensions were placed on a roller (Capco, Suffolk, UK) for 1 hour at 50 rpm. Subsequently, the suspensions were transferred to a Mettech Nanofeeder with continuous stirring.

For SPPS, GZ feedstock was prepared with the same method described previously [48], mixing gadolinium nitrate (III) hexahydrate (Fisher Scientific, Loughborough, UK) and zirconium acetate (Sigma Aldrich, Dorset, UK) in the stoichiometric proportion to form the desiered GZ phase. The ceramic yield of the solution precursor was 10 wt.%.

2.3. Topcoat Deposition

In this study, the topcoat was produced by an axial plasma spray torch (Axial III, Mettech Corp., Surrey, Canada) with a continuous feeder system (Nanofeed 350, Mettech Corp., Surrey, Canada). A plasma exit nozzle of 9.5 mm diameter was used and the feedstock was injected axially into the plasma through an injector of 1.5 mm diameter. Two variations of TBCs were obtained, which comprised the double-layered and triple-layered topcoats, as shown in Figure 1. The axial plasma gun was mounted on the robot mentioned above. The detailed spray parameters are listed in Table 1.

Table 1 Axial III spray parameters used to deposit the double- and triple-layered TBCs.

| Parameters | Suspension Yttria Stabilized Zirconia | Suspension Gadolinium Zirconate | Solution Precursor Gadolinium Zirconate |
|---|---|---|---|
| Current (A) | 200 | 180 | 200 |
| Total gas flow (L/min) | 300 | 300 | 300 |
| Ar/$N_2$/$H_2$ (%) | 44/28/28 | 44/28/28 | 45/45/10 |
| Atomising gas flow (L/min) | 20 | 20 | 10 |
| Stand-off distance(mm) | 75 | 75 | 75 |
| Scan speed (mm/s) | 1600 | 1600 | 1600 |
| Scan line distance (mm) | 5 | 5 | 4 |
| Suspension flow rate (mL/min) | 100 | 100 | 50 |
| Flame power (kW) | 120 | 121 | 122 |
| Number of passes | 25 | 30 | 40 |

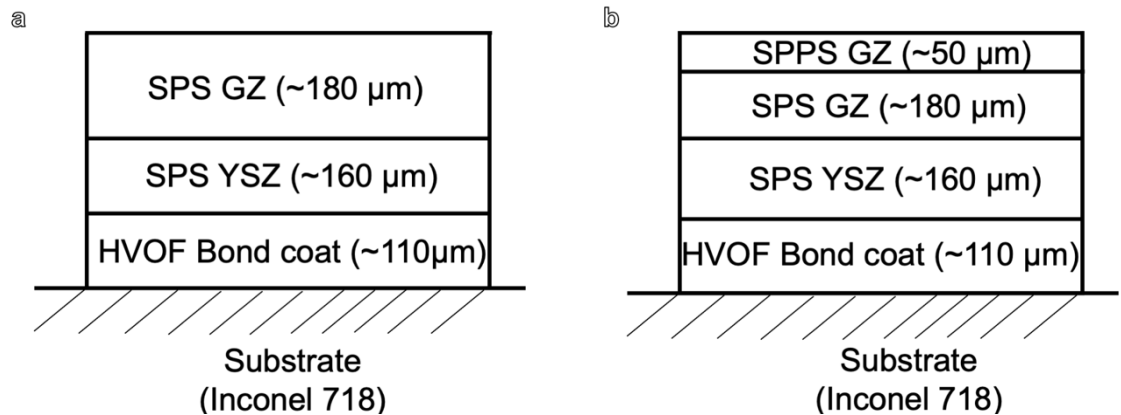

Figure 1. (a) A schematic diagram of the double-layered coating and (b) the triple-layered coating. Both coatings had the same IN718 substrate and HVOF bond-coated in the same batch. The dense layer in the triple-layered coating was achieved by plasma spraying with a solution precursor (SP) feedstock.

2.4. Material Characterisation

XRD analysis of the dried feedstock powder, the top surface of the as-sprayed and exposed TBCs was conducted using a D8 Advance DaVinci system (Bruker, Coventry, UK), equipped with a lynx eye detector. The dried feedstock powder was obtained by drying a small amount of suspension in a box furnace at 100 ºC overnight. The diffractograms were obtained with Cu-Kα radiation with a wavelength of 1.54 Å in Bragg-Brentano scanning mode. The scanning range was set from 10° to 90° 2θ, and a slow scan rate was used (0.02° step size and 0.2 s of counting time per step).

DIFFRACT.SUITE EVA software (Bruker, Coventry, UK) was used for the phase identification.

Both the as-sprayed and failed TBCs were vacuum impregnated with the epoxy resin and hardener (Struers, Rotterham, UK) at the correct proportion. The sample was then sectioned with a SiC precision cut-off wheel (MetPrep, Coventry, UK). Due to the sensitivity of the samples, a slow cutting speed (0.01 mm/s) was used. The sectioned samples were sequentially grounded with SiC grinding papers (MetPrep, Coventry, UK), and polished to 1 μm finish by diamond polishing.

For the scanning electron microscope (SEM) analysis of the as-sprayed and failed TBCs, all samples were carbon coated to obtain a conducting surface. All images were taken using Quanta 600 (FEI Europe, Netherlands) with a spot size of 50 nm, a working distance of 13 mm and an acceleration voltage of 20 kV. To understand the CMAS attack of the samples, BSE mapping was used to map the infiltrated region for Ca and Si mapping.

Besides, the coating thickness and the porosity content of TBCs were obtained with ImageJ analysis suite (NIH, Maryland, USA) [50]. The coating thickness was measured by taking the average of 10 measurements on secondary electron (SEI) images at a magnification of x150, covering approximately 1 cm of the coating cross-section with 5 images; whereas the porosity was calculated using the "analyse particle" automated function by taking the average measurement of 3 backscattered electron (BSE) images at a magnification of x300, converting the images to a black and white (8 bits) map and setting an appropriate threshold to measure the area percentage of the image covered by porosities. All measurement data is reported alongside the respective standard error.

The mechanical properties of the coating, micro-hardness, were measured using Buehler 1600 Series Micro-hardness Tester (Leinfelden-Echterdingen, Germany). A clear indentation without cracks propagated from the indented area was created with a 50-gf load and a hold time of 30 s. The reported data is calculated from an average of 12 indents with its corresponding standard error. In addition to that, a higher load in the range of 10 – 1000 gf was used to ensure cracks propagated from the edge of the indent and evaluate the fracture toughness of the samples. The fracture toughness for each layer was calculated based on the following equation developed by Evans and Wilshaw [51]:

$$K_{IC} = 0.079 \left(\frac{P}{a^{3/2}}\right) log \left(\frac{4.5a}{c}\right)$$

where $K_{IC}$ is the fracture toughness (MPa·m$^{0.5}$), $P$ is the indentation load (N), $a$ is the length of the indentation half diagonal (m) and $c$ is the crack length from the centre of the indent (m). The criteria of the criteria of $0.6 \leq c/a \leq 4.5$ must be met for the measurements to be valid.

## 2.5. Column density measurement

The columnar density was measured by drawing a horizontal line at the half of the coating thickness on a total of 10 cross-sectional SEM micrographs at x300 magnification. The total length of the cross-sectional view is approximately 1 cm. All through columns or cracks (columns or cracks from the surface of the topcoat to the bond coat interface) that intersect the line were considered. The column density was then calculated using the equation below [52]:

$$Column\ density\ \left[\frac{Columns}{mm}\right] = \frac{No.of\ columns\ intercepted\ the\ line}{True\ length\ of\ the\ line} \quad (1)$$

The variation for each measurement was considered by reporting the average value for each type of TBCs.

## 2.6. Furnace cycling test (FCT)

A total of three samples for each type of TBCs were subjected to furnace cycling tests using a programmable bottom loading isothermal furnace (CM Furnaces Inc., Bloomfield, USA). Prior to the test, all samples were heat-treated at 1135 ºC for 2 h at a slow heating and cooling rate (5 ºC/min) to burn any precursor residues that may be left in the coating. The furnace cycling test consisted of a heating stage, which heated the samples to 1135 ºC in 10 min, a dwelling stage which dwelled the samples for 45 min, and a cooling stage which cooled the samples through forced air-cooling below 100 ºC in approximately 20 min. A high-definition Webcam (Logitech C930e, Lausanne, Switzerland) was used to monitor the test, and an image was captured every 1 min interval. The test is continued until a 20% spallation area of the topcoat is observed.

## 2.7. Burner rig test (BRT)

Each type of TBCs was subjected to thermal gradient tests, also known as burner rig testing (BRT), shown in Figure 2. The BRT aims to evaluate the durability of TBCs under similar conditions to a turbine engine. Before testing, the as-deposited sample was spot welded with a Type-K thermocouple (RS Pro, Northants, UK) on the rear side (Inconel 718 surface) to monitor the substrate or back temperature. The sample was then mounted to a SS304 6 mm stainless steel tube connected to a vacuum pump system [31]. The temperature profile of the test was shown in Figure 3 alongside the temperature distribution on the front and back of the sample taken by Infrared (IR) camera (FLIR T400, Kent, UK).

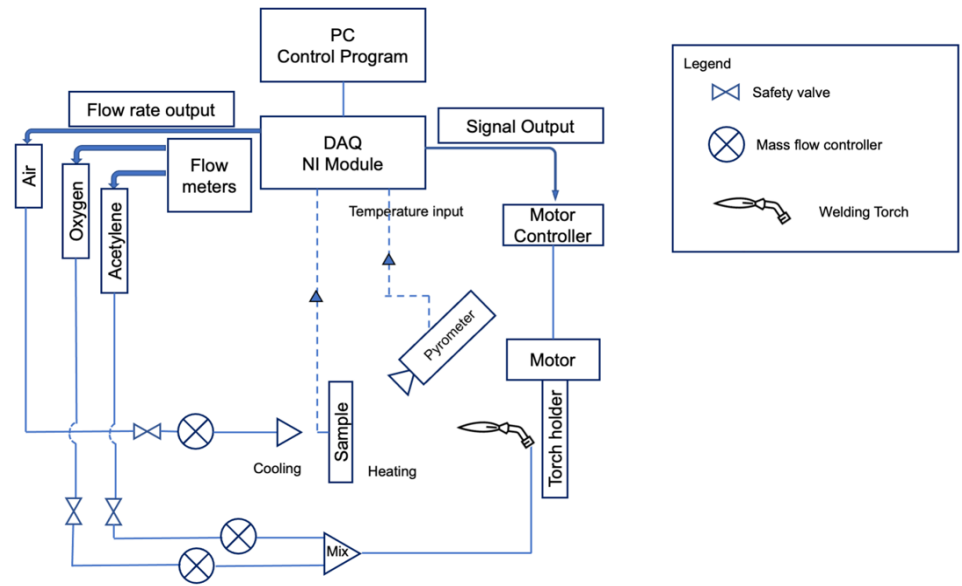

Figure 2. Schematic diagram of the BRT setup.

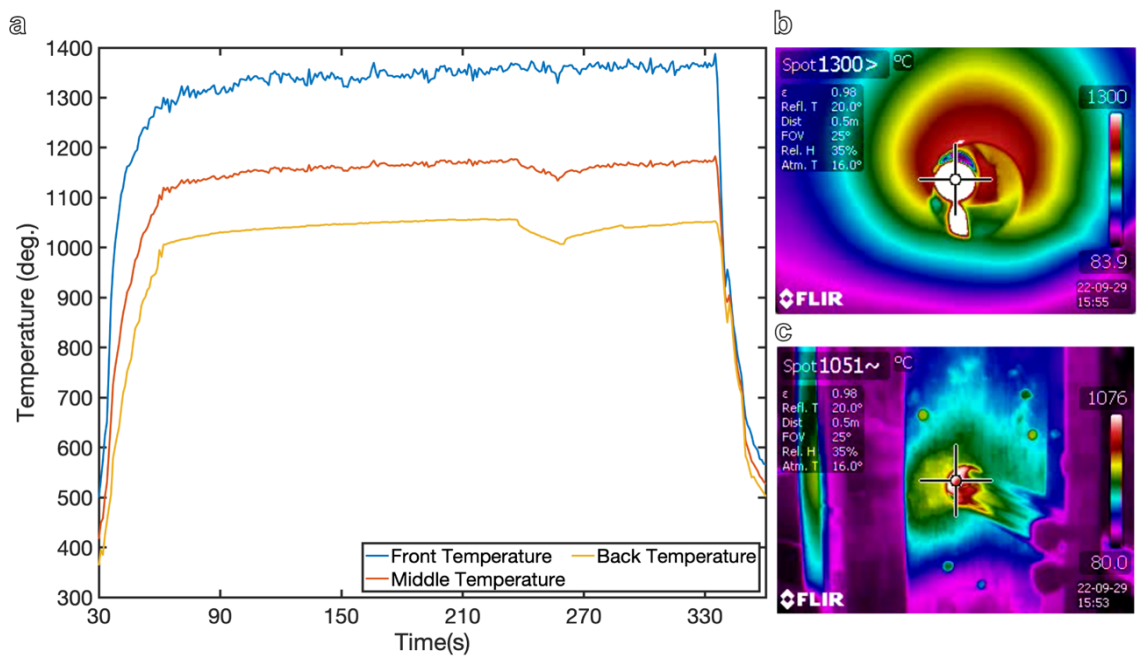

Figure 3. (a) BRT temperature profile for a cycle, which comprised of a heating cycle for 5 min and a cooling cycle for 2 min. (b) IR camera was used to measure the front and (c) back temperature distribution of the sample.

In the heating cycle, the temperature was achieved through an oxy-acetylene torch. The torch was mounted to a leadscrew and nut arm controlled by a motor controller (Igus UK Limited, Northampton, UK). Acetylene and an oxygen flow meter were used to achieve a steady oxidising flame at a temperature of 1360 °C. A single spectral pyrometer (Raytek M13, Cheshire, UK) was used to monitor the front temperature. Opposing the flame, a compressed air nozzle was directed at the back of the sample to achieve a thermal gradient across the sample. In the cooling phase, the torch was removed, and the sample was cooled from both surfaces by compressed air. The air flow rate is set high enough to cool down the sample below 100 °C within 60 to 90 s. Overall, a complete cycle comprised of 5 min of heating and 2 min of cooling. The cycle is repeated until 20% of coating spallation is observed. The middle temperature of the bond coat was then calculated using the equation below:

$$Q = kA\Delta T \qquad (2)$$

where $Q$ is the transferred heat, $k$ is the thermal conductivity, $A$ is the cross-sectional area and $\Delta T$ is the difference in temperature (K).

2.8. CMAS test

The CMAS aqueous solution was prepared by mixing the CMAS powder (Oerlikon Metco, Cheshire, UK) at a nominal composition of $35CaO-10MgO-7Al_2O_3-48SiO_2$ in mol % with deionised (DI) water at a 1:9 ratio. A uniform distribution of CMAS was then deposited on each type of the TBCs with an air brush kit. The solution was constantly agitated with a magnetic stirrer on an Isotemp hot plate (Fisher Scientific, Loughborough, UK). A CMAS concentration of 15 mg/cm$^2$ was chosen according to the previously established protocol, and guidance from the high temperature community [53]. As reported by Wellman et al. [37], an area concentration of 4.8 mg/cm$^2$ would be sufficient to cause a significant degradation against the commercial EB-PVD TBCs. Therefore, the area concentration used in this study is well above the minimum requirement. After depositing the CMAS aqueous solution, the sample was placed on a hot plate and heated to approximately 100 ºC to evaporate the DI water existed in the CMAS aqueous solution. The sample was weighted before and after the CMAS deposition. The process of depositing CMAS and sample weighting were repeated until the desired concentration was obtained.

The CMAS test was carried out in a BRF14/5 box furnace (Elite Thermal Systems Ltd., Leicester, UK). All samples were heat-treated at 1300 ºC for 5 min at a ramp rate of 10 ºC/min, with the idea to investigate how CMAS infiltrates the columnar structured TBC samples. The furnace was set to cool down to 700 ºC at the same ramp rate, 10ºC/min. Then, a slower ramp rate (5ºC/min) was used to cool down to room temperature, aiming to reduce the thermal shock behaviour that may occur in the glassy phase.

# 3. Results

## 3.1. Coating characterisation

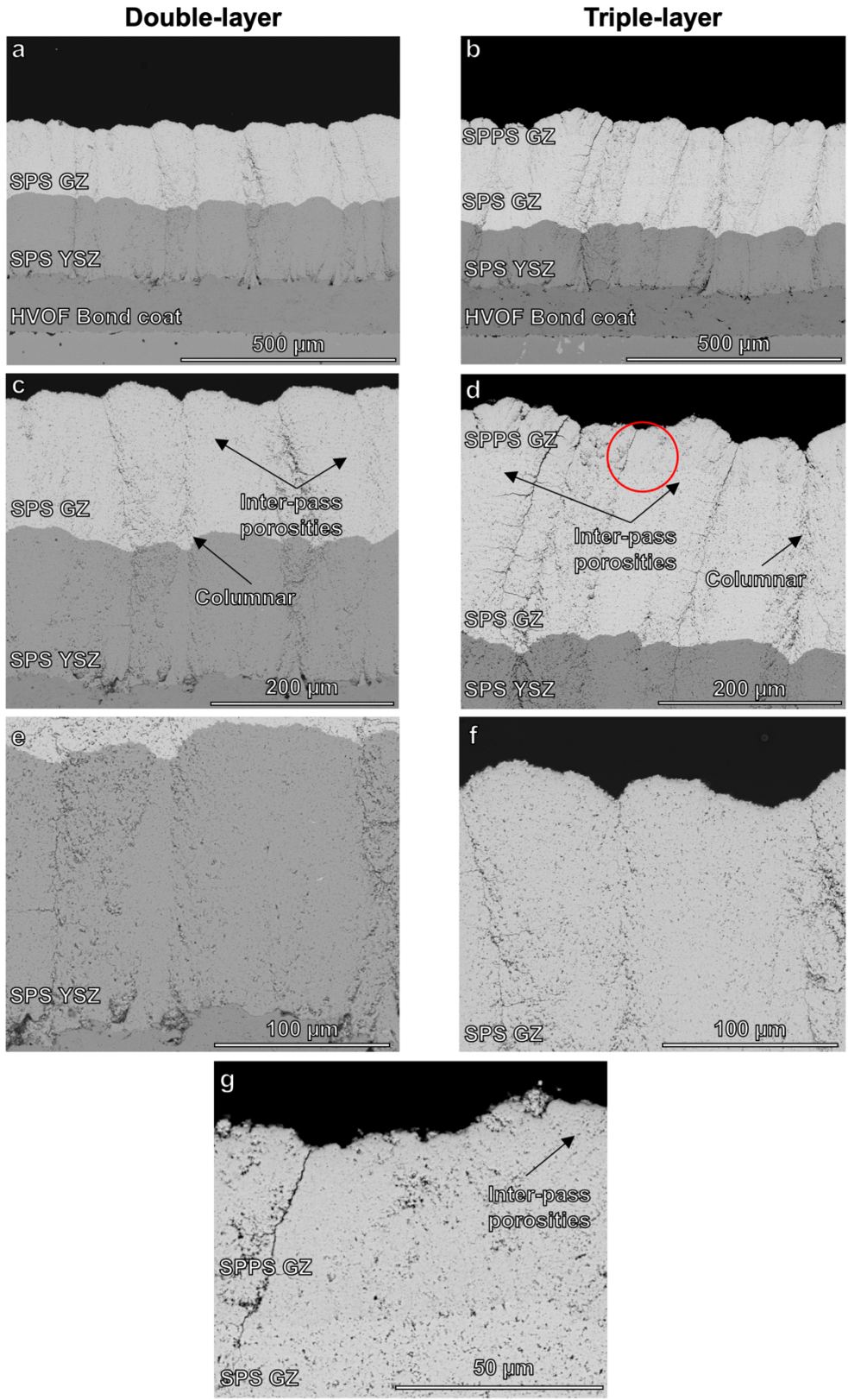

Figure 4. BSE images of (a) double-layer (SPS GZ/SPS YSZ) and (b) triple-layer (SPPS GZ/SPS GZ/ SPS SYZ) TBCs, with high magnification images shown in (c) and (d), respectively. The 'red circle' region is magnified presented in (g), showing the interface between the SPS and SPPS GZ layers. (e) High magnification of BSE images for the SPS YSZ layer, (f) SPS GZ layer and (g) SPPS GZ layer.

Two types of coatings were successfully produced, a double-layer and a triple-layer TBC. The cross-section micrographs were shown in Figure 4(a) and 4(b) respectively. Both coatings presented the typical columnar microstructure with small intercolumnar gaps, which is preferred for the TBC application as it offers good strain-tolerance capabilities; The dense SPPS GZ layer (i.e., triple-layer coating as shown in Figure 4(b), 4(d) and 4(g)) exhibits a columnar feature that is intermediate between the vertical crack and columnar microstructures. A similar intermediary microstructure is also observed in other works, reported by Ganvir et al. and Curry et al. [28,54,55]. Interestingly, most of the columnar gaps present in the SPS GZ layer were found to be minimised in the dense SPPS GZ layer due to a finer splat size attributed to the SP feedstock. As mentioned in Section 2.1, all samples were HVOF thermal sprayed bond coat with a thickness of 117 ± 2 μm in the same batch to avoid any run-to-run variation.

The main difference between the two coatings is the additional dense SPPS GZ layer, resulting in an addition of approximately 55 μm to the total thickness for the triple-layer coating. The double-layer SPS GZ/SPS YSZ TBC had a total thickness of 335 ± 7 μm while the triple-layer SPPS GZ/SPS GZ/SPS YSZ TBC had a total thickness of 398 ± 5 μm. Since both the SPS YSZ and SPS GZ layers were also deposited in the same batch for either the double-layer or the triple-layer TBCs, the SPS YSZ layer had an average thickness of 152 ± 4 μm while the SPS GZ layer had an average thickness of 178 ± 6 μm; The dense SPPS GZ layer had an average thickness of 55 ± 5 μm.

On the other hand, column densities in columnar microstructures are found to be an important factor for a long lifetime in thermal cycling tests as the strain-tolerance capability of the coating structure is improved [56]. Based on Figure 4(a) and 4(b), the column density of the double-layer coating was found to be 9 ± 1 columns/mm and the triple-layer coating was found to be 10 ± 2 columns/mm. However, it is worth to be noted that the reported average value was calculated by only taking into consideration of through columns or cracks, as mentioned in Section 2.5. Comparing these values with other works, the reported column densities lie within the acceptable range of 7 – 9 columns/mm [57]. In terms of the mechanical properties of each layer, the SPS YSZ had the highest microhardness and fracture toughness values while the SPS GZ and SPPS GZ had a similar value, as shown in Table 2. The measured values agreed with the literature where GZ tends to have a lower fracture toughness value than YSZ [20,24,25,58,59];

However, the reported hardness value for the SPS YSZ is higher than the literature ones. A possible explanation for that is the porosity level of SPS YSZ in the literature ones is higher (~ 20%), hence a lower hardness value is expected. Moreover, it is also worth stating that the fracture toughness measured by micro-indentation method may not represent the overall fracture toughness of the TBC as it represents the localised values only. Intercolumnar gaps and inter-pass porosities bands cannot be reliably investigated with a micro-indenter, hence an additional technique is required to measure the fracture toughness of the entire TBC.

In the meantime, porosities in the coating structure also play an important role in the lifetime of TBCs as a high porosity coating structure favours lower thermal conductivity; however, it would reduce the fracture toughness of the coating structure [58]. Therefore, it is essential to evaluate the porosity of the as-sprayed TBCs. The measured porosity in the individual layers for both coatings (i.e. double-layer and triple-layer), SPS YSZ, SPS and SPPS GZ, was reported in Table 2. Although YSZ had a higher melting point (2700 °C) than GZ (2570 °C), both SPS YSZ and SPS GZ layers had the same porosity level [19]. It could be explained by the lower current utilised for depositing SPS GZ layer while all other parameters were kept constant, hence GZ splats are expected to have the same degree of melting with YSZ; The SPPS GZ layer had the lowest porosity, resulting in a relatively denser structure than the SPS YSZ and the SPS GZ layers. The dense structure can be said to have successfully achieved through the SPPS deposition method.

In summary, the total porosity in SPS and SPPS coating structures is mainly contributed by the columnar features (known as intercolumnar porosities) and the layered porosities present in the coating structure (known as inter-pass porosities (IPBs)) [26,60]. However, it is worth mentioning that the image analysis method may not measure the fine-scaled porosities (in the range of nanometres) in the SPS and SPPS coating structures at x300 magnifications. Thus, an additional method would be required to accurately predict the actual porosities in the coating structures, considering the submicron to nanometre range porosities along with open and closed porosities.

Table 2 Mechanical properties and porosities of each layer for both double- and triple-layer coatings.

| Layer | Microhardness, $HV_{0.5}$ | Fracture toughness, $MPa.m^{0.5}$ | Porosities (%) |
|---|---|---|---|
| SPPS GZ | 662 ± 15 Hv | 0.58 ± 0.3 | 8 ± 2 |
| SPS GZ | 658 ± 29 Hv | 0.46 ± 0.4 | 10 ± 1 |
| SPS YSZ | 714 ± 32 Hv | 1.91 ± 0.2 | 11 ± 1 |

*3.2.X-ray diffraction of coatings*

A series of XRD diffractograms consisting of the GZ dried powder from suspension, top surface of the double-layer and triple layer coatings was presented in Figure 5. Comparing the diffractogram of the S-GZ dried powder with the as-sprayed coatings (i.e., double- and triple-layer coatings), all main peaks represented the cubic fluorite GZ (PDF Card #080-0471); however, a small amount of pyrochlore GZ (PDF Card #080-0470) was detected in the as-received GZ suspension. The absence of the pyrochlore GZ in both as-sprayed coatings indicated that the GZ particles were completely molten during the deposition process and the rapid solidification of molten splats suppressed the crystallisation of pyrochlore phases. Additionally, RE zirconates undergo an order-disorder transition at high temperatures, typically above 1500 ºC, from a pyrochlore structure to a cubic fluorite structure after holding at this temperature for ~10 h or more [61–63]. Thus, the formation of the GZ pyrochlore structure is not possible to occur in the deposition process as the deposited splats will undergo rapid cooling. A similar finding is also observed by Bakan et al. [64]. In the meantime, the peaks of the as-sprayed triple-layer coating (i.e., SPPS GZ) show no difference to the as-sprayed double-layer coating (i.e., SPS GZ). A more detailed work of the SP-HVOF thermal spray of GZ feedstock was studied in [48].

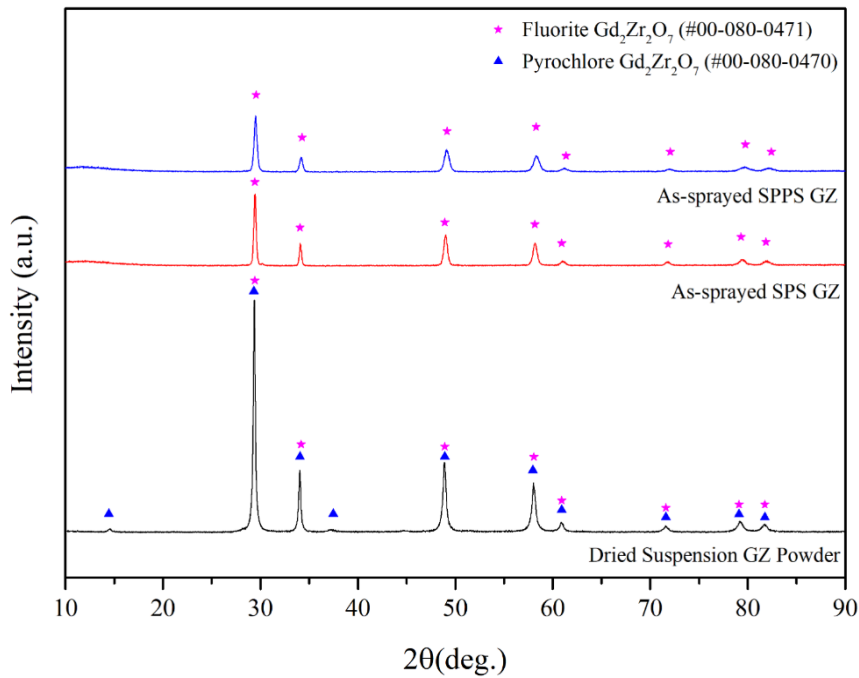

Figure 5. XRD peaks of the dried SGZ powder, the as-sprayed double-layer (SPS GZ) and triple-layer coatings (SPPS GZ).

## 3.3. Furnace Cycling Test (FCT)

Both types of samples (i.e., double- and triple-layer coatings) were subjected to cyclic tests at 1135 ºC in a bottom-loading furnace and compared. It is worth stating that there is no temperature gradient across the samples. As shown in Figure 6a, the triple-layer coating presented a similar thermal cycling lifetime of 97 % in relative to that of the double-layer coating.

The failed samples were cross-sectioned and investigated through SEM micrographs, as shown in Figure 7. Both coatings failed at the TGO/topcoat interface, possibly due to the stress accumulation induced by the thickening of the TGO layer and CTE mismatch between the ceramic topcoat and the substrate. A similar failure mode is also reported previously on multi-layer TBCs being subjected to furnace cycling tests [25,65]. The measured TGO was found to be 6 ± 1 μm, in which the TGO critical thickness was reported to be in the range of 5 – 8 μm depending on the composition of the bond coat [65,66]. The TGO critical thickness is defined as the maximum thickness that the TGO can grow before the spallation of TBC occurs. In this case, the TGO layer is found to exceed the critical thickness, resulting in the spallation of TBCs and causing failure.

The high magnification images in Figure 8 reveal that some of the intercolumnar gaps have widened and the nucleation of vertical cracks has begun within the columnar (Figure

8b). Additionally, the inter-pass porosities in the as-sprayed TBCs were also significantly reduced, which can be mainly attributed to the sintering of the topcoat during the heating cycle of the cyclic tests. As a result, the sintering process stiffens the topcoat, leading to the generation and propagation of vertical and horizontal cracks in the coating structure [67]. Interestingly, Table 3 shows that the microhardness value of each layer was marginally higher than the as-deposited condition, except for the SPPS GZ layer. The microhardness value for the SPPS GZ layer cannot be accurately determined due to several factors, including the layer being too thin, and the presence of micro-scaled porosities and cracks within the structure. Moreover, it is also observed that the horizontal cracks propagated laterally to the adjacent crack or column along the inter-pass porosities as voids or microcracks are less resistant pathways, easing the propagation of cracks when the accumulated strain energy exceeds the fracture toughness of the respective layer.

Table 3 Microhardness value of the as-sprayed sample, FCT exposed-, BRT exposed- and CMAS-exposed samples, respectively.

| Layer | Microhardness, $HV_{0.5}$ | | | |
| --- | --- | --- | --- | --- |
| | As-sprayed ($HV_{0.5}$) | FCT ($HV_{0.5}$) | BRT ($HV_{0.5}$) | CMAS ($HV_{0.5}$) |
| SPPS GZ | 662 ± 15 Hv | 566 ± 47 Hv | 655 ± 22 Hv | 718 ± 20 Hv |
| SPS GZ | 658 ± 29 Hv | 676 ± 22 Hv | 757 ± 5 Hv | 760 ± 29 Hv |
| SPS YSZ | 714 ± 32 Hv | 750 ± 20 Hv | 842 ± 18 Hv | 927 ± 16 Hv |

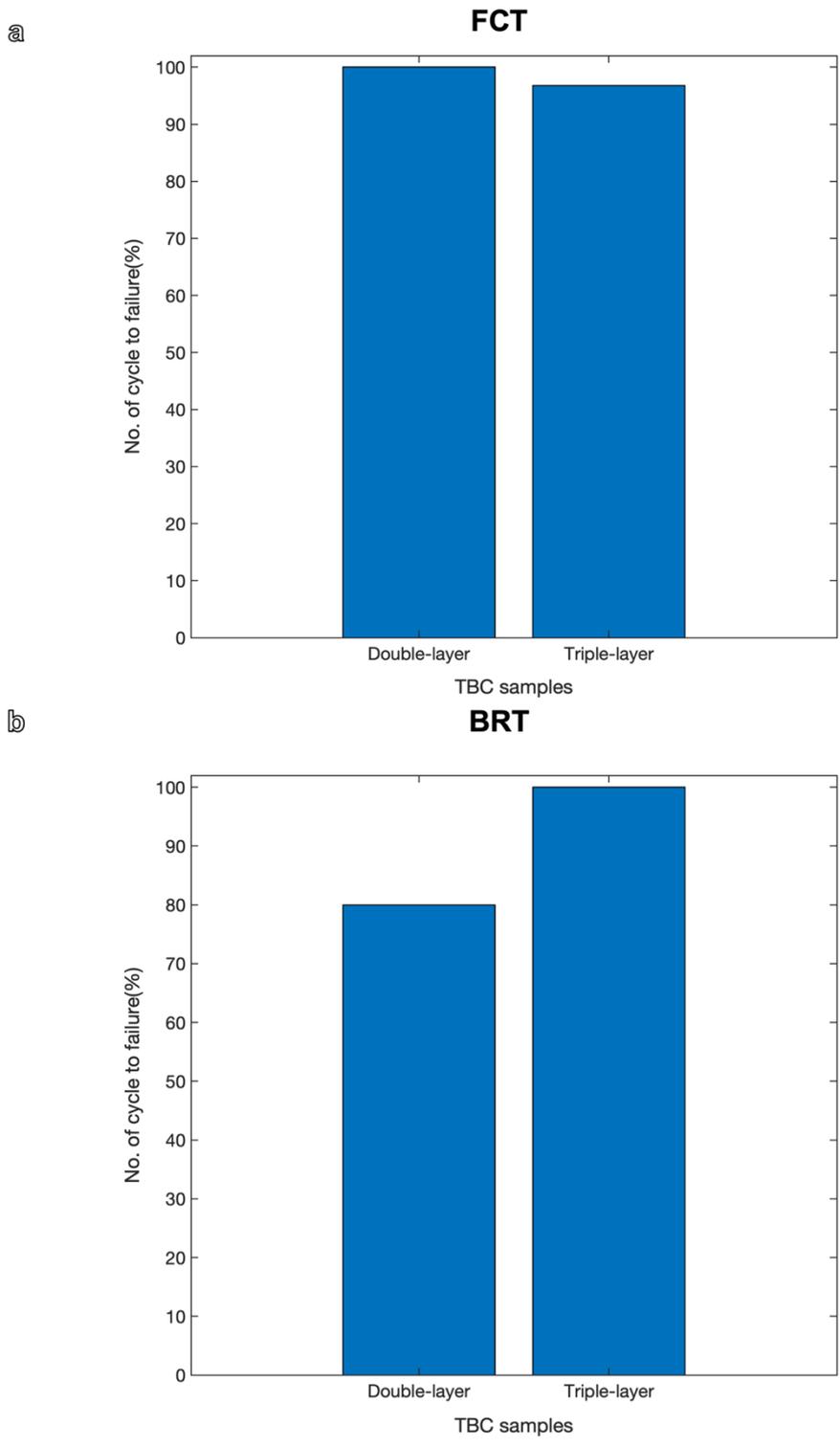

Figure 6. Lifecycle of both double-layer and triple-layer coatings in (a) furnace cyclic test (FCT) and (b) burner rig testing (BRT). The plots represented the lifecycle of the triple-layer coating in relative to the double-layer coating, treating the double-layer coating as the baseline sample.

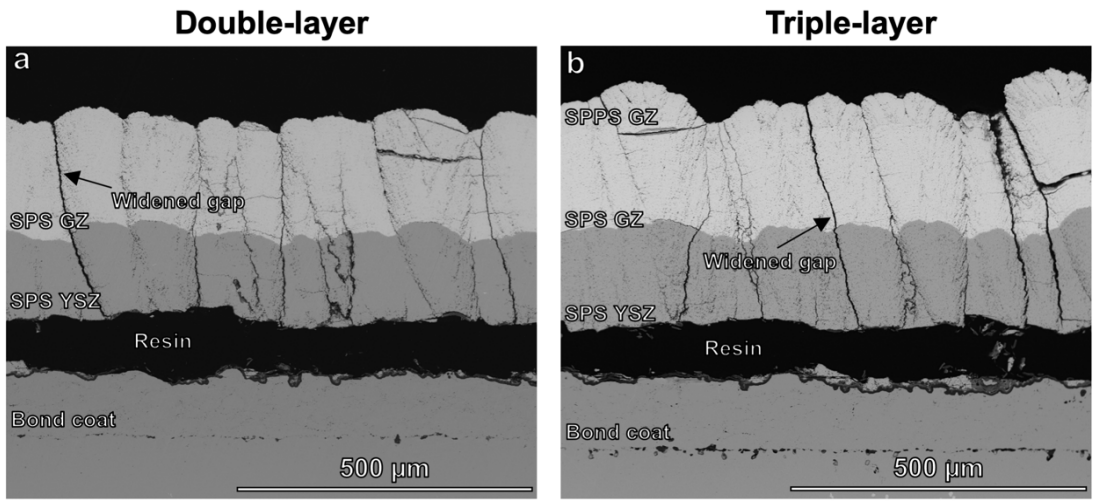

Figure 7. SEM micrographs of the FCT failed samples, (a) the double-layer coating and (b) the triple-layer coating.

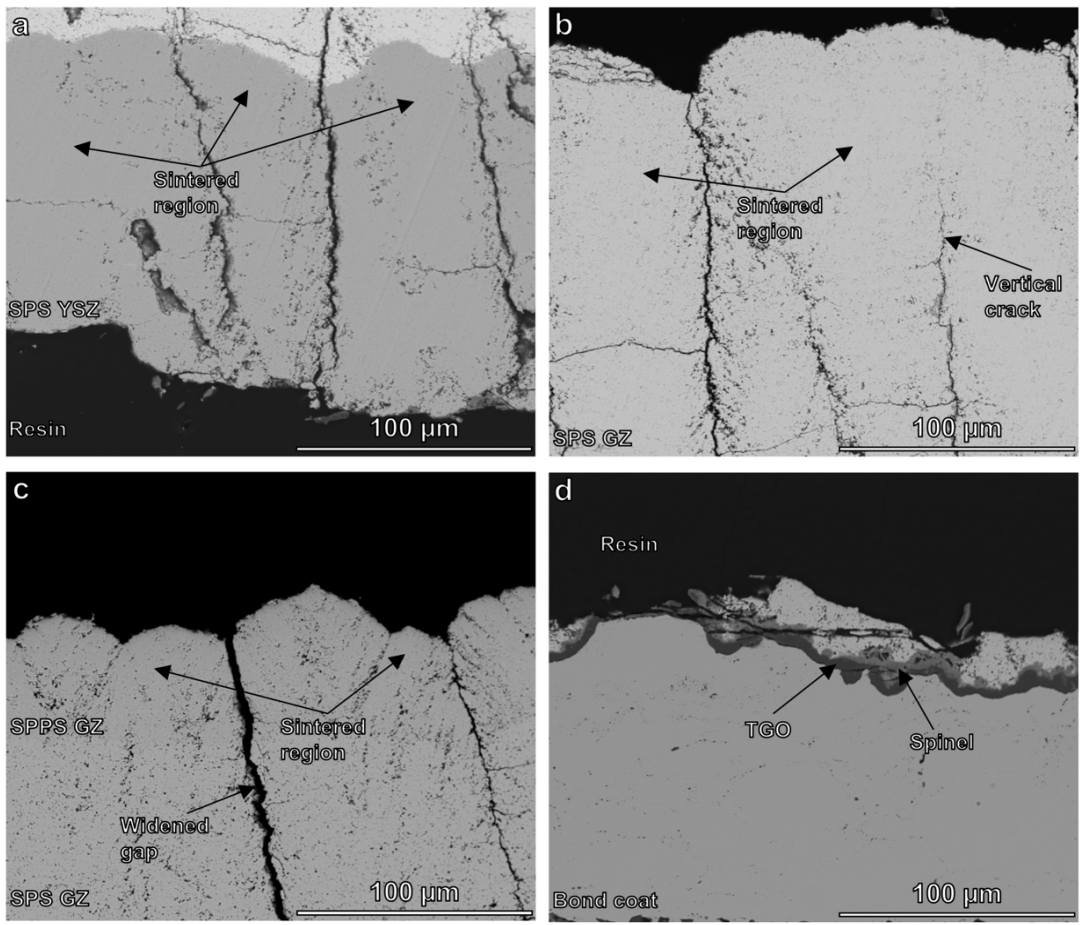

Figure 8 High magnification SEM micrographs of the FCT failed samples for each individual layer, (a) SPS YSZ (b) SPS GZ (c) SPPS GZ and (d) Bond coat. Inter-pass porosities in the as-sprayed coating are no longer to be seen in some regions of the coating while the bond coat is fully depleted in β-phase.

### 3.4. Burner Rig Test

The lifetime of both coatings was also investigated in thermal gradient tests (i.e., BRT tests), as shown in Figure 9. The triple-layer coating demonstrated a higher lifecycle compared to the double-layer coating, with the latter having approximately 20 % lower lifecycle, as illustrated in Figure 6b. The digital photographs of the exposed samples were depicted in Figure 9a and 9b, where the region with dark blue appearances suggested the spallation of the topcoat. It should be noted that spallation occurred in large chunks, making it difficult to control the spallation area within 20 %. Samples were stopped from cycling if spallation exceeded the defined threshold.

The failed samples were cross-sectioned and investigated through SEM micrographs as shown in Figure 9(c) – (e). Figure 9c and 9d presented the partial failure mode that occurred at a particular region for both coatings (double- and triple-layer coatings), where the SPS and SPPS GZ layers were delaminated layer-by-layer through the top of the SPS YSZ layer. It can be observed that lateral cracks were initiated and propagated along the inter-pass porosities present in the coating structure, interlinking with the adjacent columns or cracks, causing the subsequent layers to delaminate after achieving a number of cycles. Figure 9(e) and 9(f) showed the complete failure mode of the topcoat in both samples, where the topcoat delaminated from the topcoat/TGO interface. A possible reason could be due to the CTE mismatch between the topcoat and the substrate. According to the higher magnification SEM micrographs taken at regions where the individual layer in the topcoat was still intact (Figure 10(a) and Figure 10(b)), the inter-pass porosities that already present in the coating structure were significantly reduced. The significant reduction in pores suggested that the TBCs experienced a higher sintering effect in BRT than FCT, as evidenced by the compelling increment in the microhardness value for the BRT exposed sample (Table 3). However, it is worth noting that the SPPS GZ layer showed a similar microhardness value to the as-deposited condition, which could be explained that the layer was being too thin and requiring other methods to improve accuracy.

Meanwhile, the TGO layer is observed to be relatively thin as compared to the samples exposed in FCTs (Figure 10(c)). The measured TGO thickness for both coatings is $1.5 \pm 0.5$ μm, which is lower from the critical TGO thickness of 5 – 6 μm, indicating that the TGO is not the dominant for the failure mode. The thin TGO layer could be due to the rapid heating cycle in BRT (5 mins of heating and 2 mins of cooling) as compared to FCT, where the samples were held at the testing temperature (1135 °C) for 45 mins in each cycle. A similar finding is also observed from the previous BRT tests on APS and SPS TBCs [25,64].

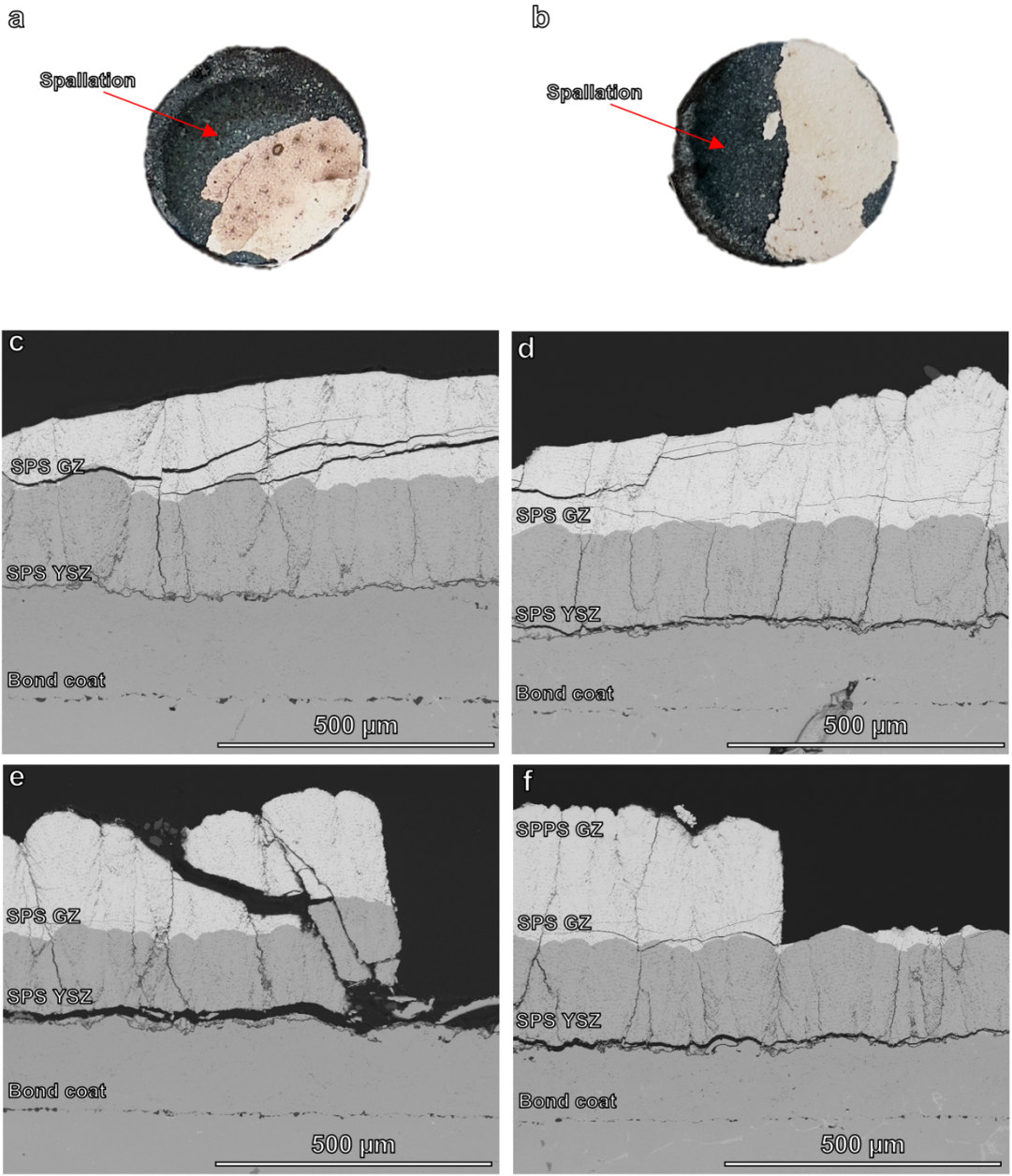

Figure 9. (a) Photographs of the failed double-layer and (b) triple-layer coatings. SEM micrographs show the early failure mode of the (c) double-layer and (d) triple-layer coatings. (e) SEM micrographs show a complete failure mode of the failed double-layer and (f) triple-layer coatings.

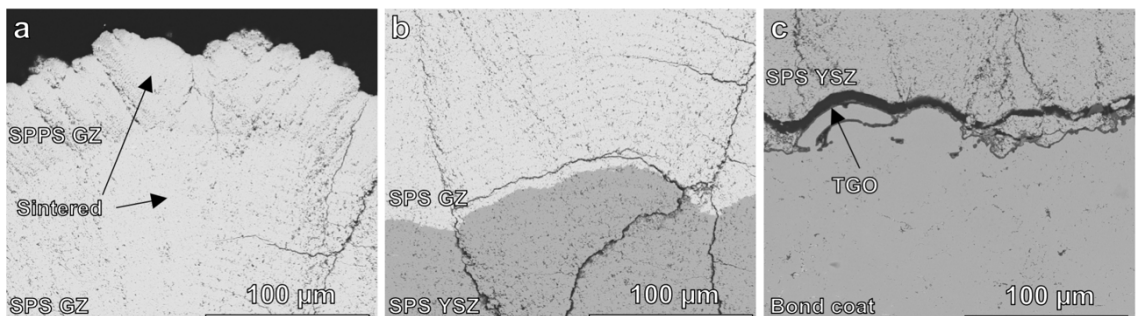

Figure 10. High magnification of SEM micrographs of the exposed coatings after BRT tests at (a) SPS GZ and SPPS GZ interface, (b) SPS GZ and SPS YSZ interface and (c) the bond coat. These micrographs were taken at regions where the individual layer in the topcoat was still intact to each other.

*3.5. CMAS test*

The CMAS composition used in this study had a glass transition temperature of ~ 800 ºC and a melting temperature of ~ 1220 ºC, thus the CMAS is expected to be fully molten at the testing temperature [53]. As evidenced in Table 3, both TBCs (i.e., double- and triple-layer coating), exhibited a significant increase in microhardness values after exposed to CMAS. This observation is consistent with the findings reported by Lokachari et al. [68].

For the double-layer coating, the topcoat (SPS GZ/SPS YSZ) is fully infiltrated by CMAS. It is observed that the spallation occurred at the SPS GZ/SPS YSZ interface, as shown in Figures 11(a) and 11(b). Horizontal cracks were seen to propagate along the inter-pass porosities in the SPS GZ layer and the regions adjacent to or at the SPS GZ/SPS YSZ interface. It can be explained by the low fracture toughness of the SPS GZ, where the stress level in the coating structure exceeds the fracture toughness of GZ after infiltrating by CMAS [69,70]. Since the fracture toughness for the SPS YSZ is higher, the SPS YSZ layer resisted delamination, thus horizontal cracks mainly propagate locally in the SPS GZ layer or at regions near to the interface between SPS GZ and SPS YSZ. Additionally, horizontal cracks were also seen to propagate in the SPS YSZ layer, specifically near to the topcoat/TGO interface, which could be attributed to the CTE mismatch between the topcoat and the substrate. From the EDX mapping in Figure 11(c) – 11(f), a high contrast of Ca and Si maps were detected along the wide-opened columnar gaps in SPS GZ, suggesting that the CMAS infiltrated the coating through these gaps to the SPS GZ/SPS YSZ interface. The initiation and propagation of vertical cracks are also observed in the SPS GZ layers. From the EDX mapping shown in Figure 11(d) and 11(f), these vertical cracks could be induced by CMAS infiltration or due to the sintering of the topcoat. It is worth mentioning that the SEM micrographs were taken after the samples failed. Thus, a further investigation would be required to determine the dominant factor that leads to the formation of these vertical cracks.

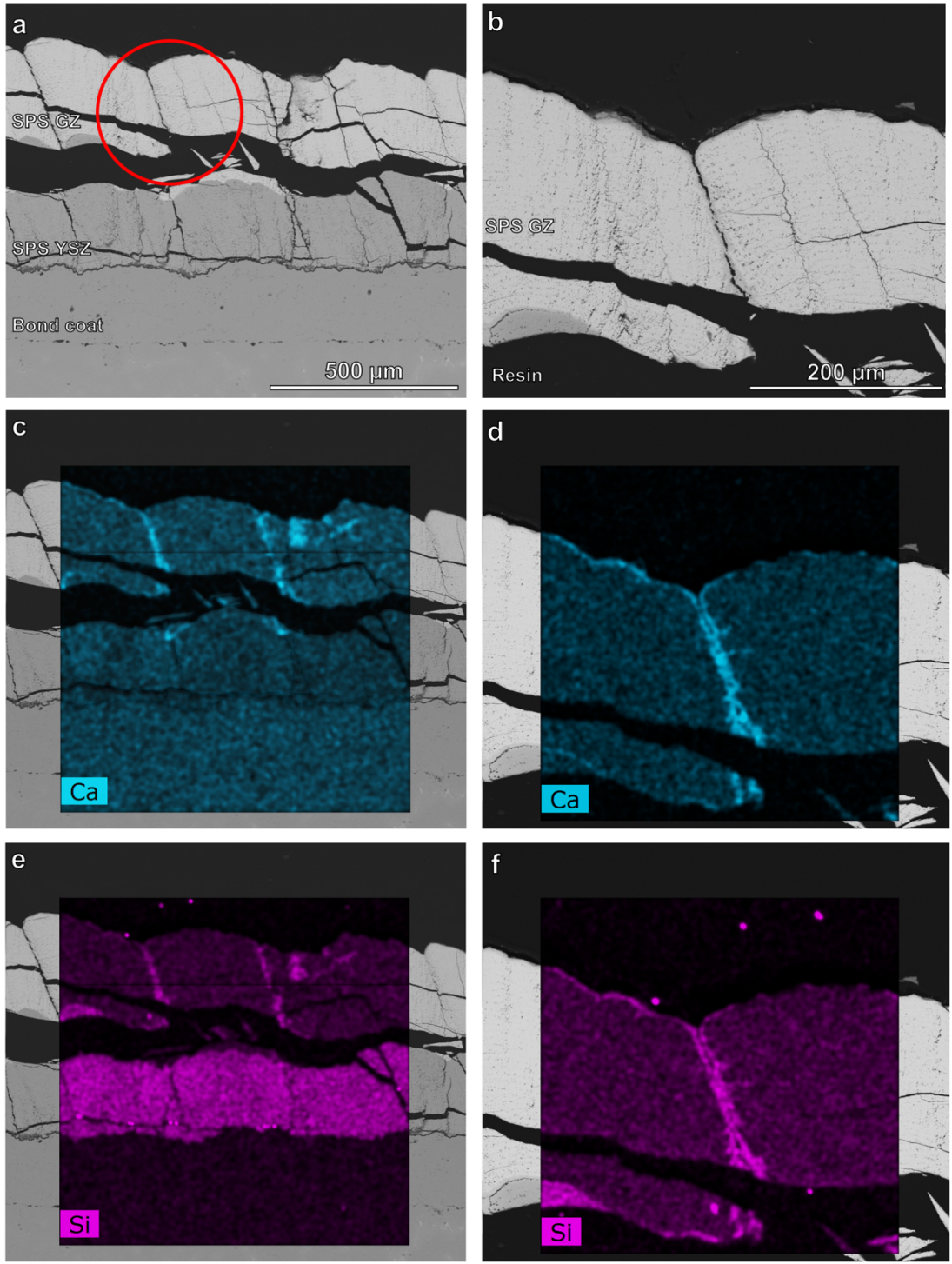

Figure 11. (a) SEM micrograph of the CMAS tested double-layer sample. (b) High magnification image of the 'red circled' region. (c) &(d) EDX mapping for Ca element (d) EDX mapping for high magnification image; (e) & (f) represents the EDX mapping for Si element (f) EDX mapping for the high magnification image.

For the triple-layer coating, the topcoat is fully infiltrated along the through channels, which are defined as the columnar gaps that developed from the SPS YSZ layer to the top surface of the topcoat (SPPS GZ layer). The topcoat is partially infiltrated along the minimised columnar features or cracks and stopped at the SPPS GZ/SPS GZ interface.

The infiltration depth of these partial infiltrated columns or cracks is measured to be within the range of 38 μm to 68 μm. According to Figures 12(a) and 12(b), one spallation occurred at the SPS GZ/ SPS YSZ interface (a similar spallation to the double-layer coating) while the second spallation occurred at the region close to the SPS YSZ/TGO interface. The reason for the latter spallation is possibly due to the higher total thickness in the triple-layer coating, resulting in higher residual stress in the coating structure as the residual stress is in a linear proportion to the coating thickness [71]. After infiltrating by CMAS, the volume changes due to the phase transformation in the SPS YSZ layer and the CTE mismatch between topcoat and substrate would increase the stress level in the coating further, inducing spallation in the SPS YSZ layer at regions near the TGO [69]. Based on the EDX mapping in Figure 12(c) – (f), the traces of CMAS are found along the intercolumnar gaps, indicated by the high contrast of Ca and Si maps. Interestingly, the width of the infiltrated through channels is measured to be 4.6 ± 1.5 μm, while the width of the partially infiltrated channel is measured to be 0.6 ± 0.3 μm. Although Kumar et al. reported that the DVC cannot be sealed if the width is wider than 1 μm, further investigation is required to justify if the partially infiltrated channels were sealed as a result of the reactant product with CMAS or the infiltration was not completed due to the short exposure time (5 min dwell time at 1300 °C) [29]. Furthermore, cracks induced by CMAS are also observed at the top surface of the coating (shown in Figure 12(g)). The reason is possibly due to the low thermal expansion and high hardness of CMAS, inhibiting the coating from contracting in the cooling stage, resulting in increased shear and tensile stresses in the coating structure [69,70].

By comparing the CMAS residue on the top surface of the double-layer (Figure 11(d) and 11(f)) and the triple-layer coatings (Figure 12(d) and 12(f)) through Ca and Si maps, a higher contrast was detected and observed in the triple-layer coating, showing that there was more CMAS residue in the triple-layer coating after exposing at 1300 °C for the same amount of time (5 min). A possible explanation is that the triple-layer coating has lesser through channels that can possibly guide the molten CMAS to infiltrate the entire coating as compared to the double-layer coating. Therefore, the molten CMAS has to find an alternative route to infiltrate the entire topcoat, thereby decreasing the infiltration speed in the triple-layer coating.

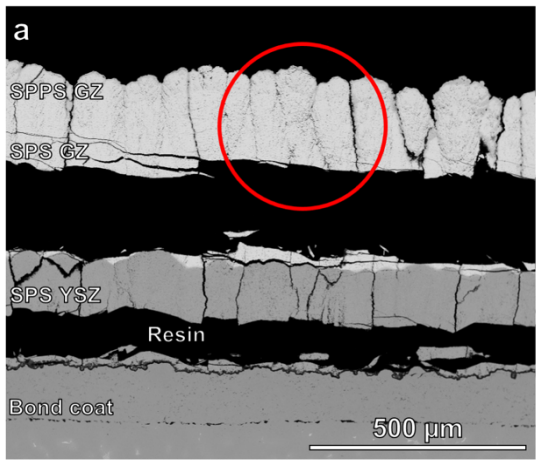
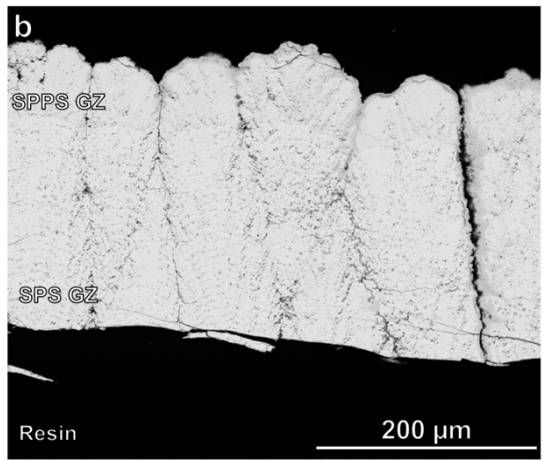
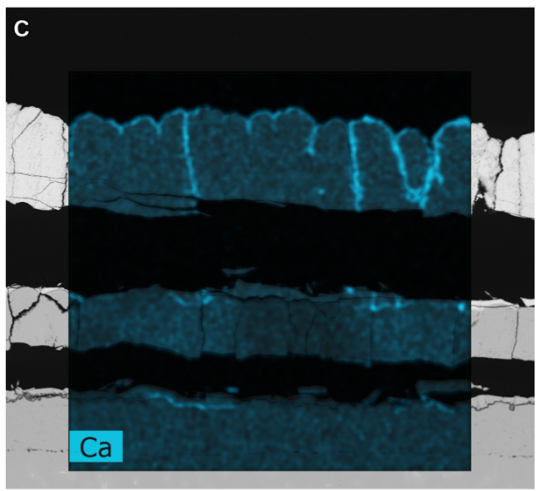
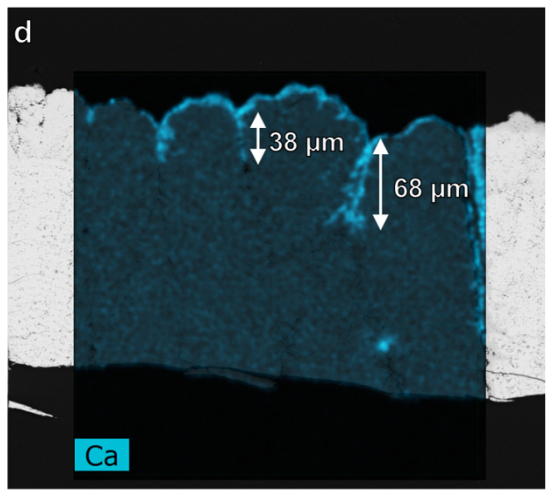
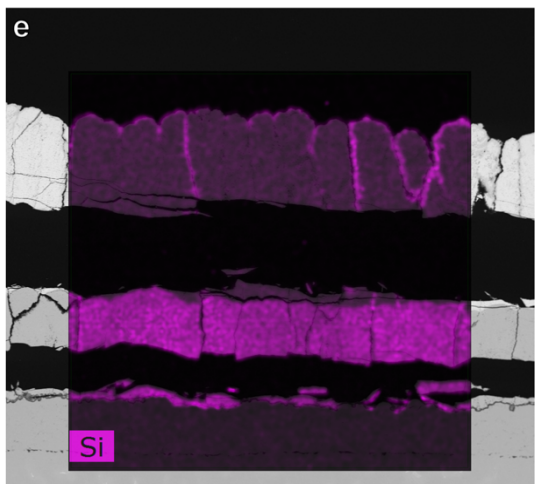
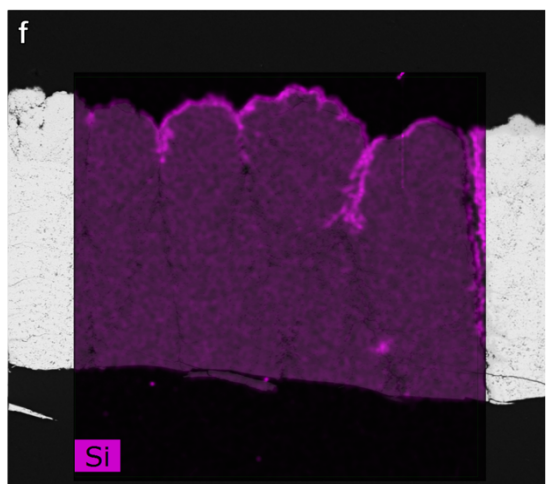
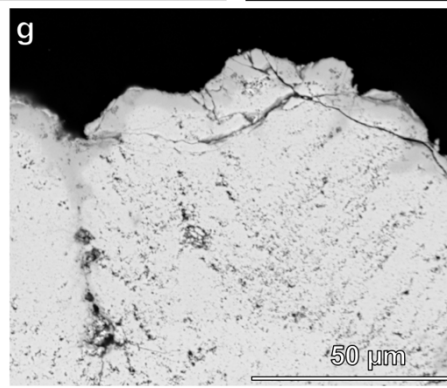

Figure 12. (a) SEM micrograph of the CMAS tested triple-layer sample. (b) High magnification image of the 'red circled' region. (c) &(d) EDX mapping for Ca element (d) EDX mapping for high magnification image; (e) & (f) represents the EDX mapping for Si element (f) EDX mapping for the high magnification image. (g) High magnification at the top surface of topcoat, SPPS GZ layer, where cracks were induced by CMAS upon cooling.

## 4. Discussion

### 4.1. Microstructure of the as-sprayed coating

Both as-sprayed coatings (double- and triple-layer) presented the typical SPS columnar structure that is proven to offer good strain-tolerant capabilities and lower thermal conductivities [25,55,57]. Interestingly, the columnar formation is mainly due to the redirection of the plasma plume after contacting the substrate and forming an adherent boundary layer [72,73]. In the deposition process, the spray direction is almost perpendicular to the surface of the substrate, but the plasma jet is deflected parallel to the surface of the substrate as the plasma jet impacts the substrate. As a result, the direction and the flow velocity of the plasma flow encountered a significant change near the substrate, from normal to parallel to the surface of the substrate, resulting in a plasma drag force being exerted on the in-flight particles [73,74].

According to Table 1, the parameters used to deposit the SPS YSZ and SPS GZ layers are relatively high in total gas flow and atomising flow rate, resulting in a strong atomising effect on the injected feedstocks. The suspension droplets break up into relatively small droplets, the solvent surrounding the droplet evaporates, and the solid particles form small agglomerates which are then melted and accelerated. Most of these in-flight particles that travel in the plasma trajectory tend to have a low momentum and inertia and follow the plasma flow adjacent to the surface of the substrate, depositing on the side of surface asperities or being unable to follow the sudden change in the plasma flow [73,74]. The successfully deposited TBC materials on the side of surface asperities creates an inter-deposit gap. With the continuation of the deposition process, the growth of the deposits continues, and the inter-deposit gap produces the columnar structure across the coating thickness [74]. In addition to that, IPBs are also observed in both the coating structures. It is worth noting that the utilised suspension feed rate is relatively high, 100 mL/min. Thus, the formation of IPBs suggested that partial of the in-flight particles are more likely to be entrained in the plume periphery instead of the plasma core. The in-flight particles in this region are usually slower and cooler, resulting in a semi-molten or re-solidified particle, traveling along with the fully molten splats. Due to the repeated transverse pattern of the plasma torch which corresponds to the number of passes, these semi-molten or re-solidified particles deposited in between the molten splats, leading to layers of porosities between successive passes [55,57]; However, IPBs can be detrimental to the lifetime of the coatings as it provides an easy pathway for the propagation of horizontal cracks.

For the deposition of the SP feedstock, the deposition mechanism differs from the suspension feedstock. After injecting into the plasma and breaking up into droplets, the droplet will undergo evaporation, droplet breakup, precursor decomposition and sintering, heating and melting of the solid particles. Since the SP feedstock is axially injected into the plasma plume, the injected droplets are expected to have a better entrainment and be more effective in heating and melting before depositing onto the substrates [75]. Referring to Figure 4(b) and Figure 4(d), the deposition of the dense SPPS GZ layer succeeded in continuing and minimising the intercolumnar gaps from the SPS GZ layer. The minimisation of these intercolumnar gaps is mainly attributed to the finer splat size produced by the SP feedstock [76,77]. Apart from columnar features, vertical cracks can also be seen in the dense SPPS GZ layer, presenting an intermediatory structure between vertical crack and columnar structures. The formation of these vertical cracks could be due to the high residual stress accumulated in the coating structure and the un-melted particles incorporated within the splats [75,77]. The high residual stress in the coating structure acts as the driving force for the initiation and propagation of the vertical cracks in the coating structure, especially at regions where multiple un-melted particles are deposited in close vicinity.

## 4.2. Failure mode between FCT and BRT

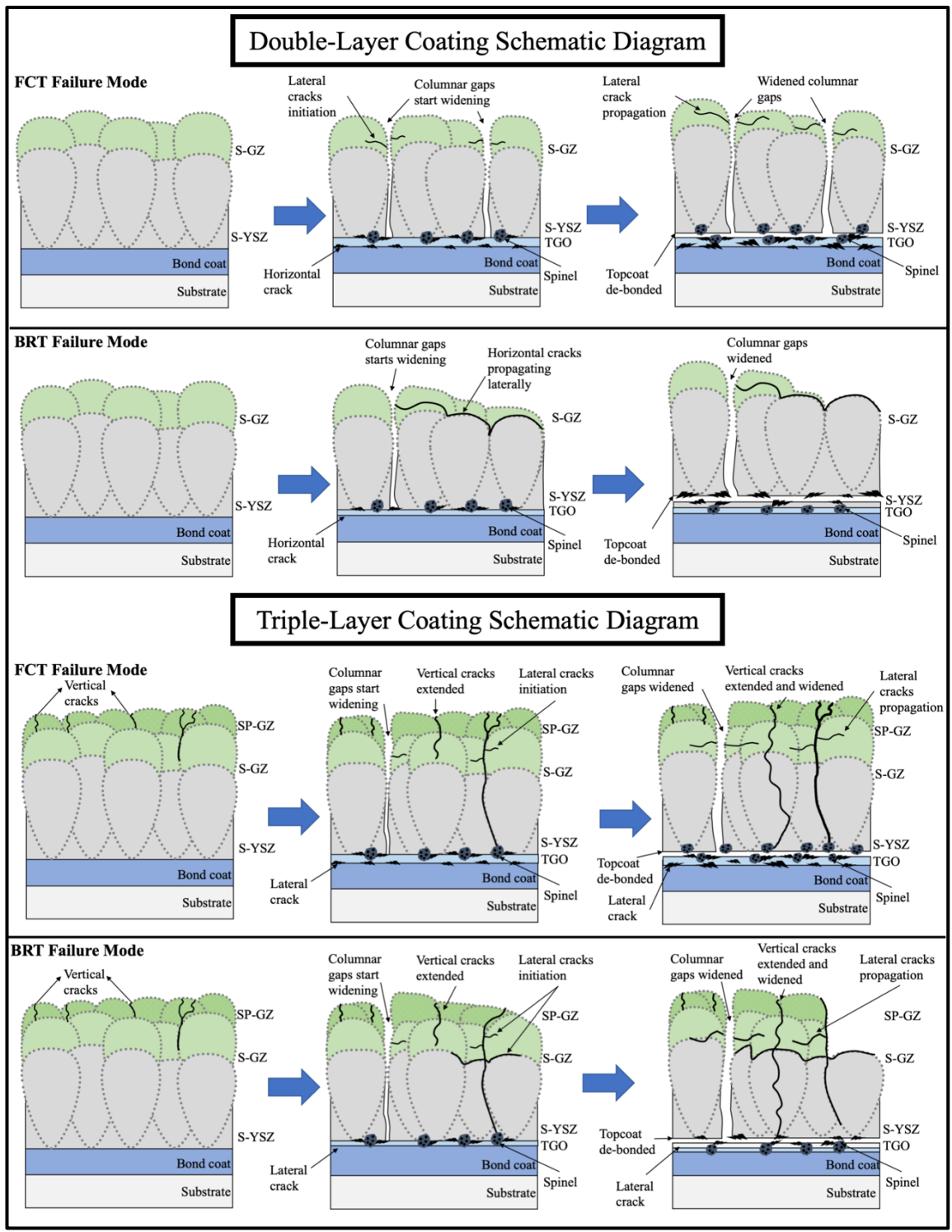

Figure 13. Schematic diagrams for the predicted failure mode in the double-layer and the triple-layer coating. A comparison is made between the FCT and BRT test condition, leading to a different failure mode in each type of the coating.

The two types of coatings (i.e., double-layer and triple layer TBC) were subjected to both FCT and BRT. The main difference between the FCT and BRT is the test condition, where there is no temperature gradient across the sample in the FCT. Thus, the failure mode of the samples varied between the FCT and the BRT (as shown in Figure 7 and Figure 9). In the FCT (Figure 6a), both the coatings had the same lifetime, whereas in the BRT (Figure 6b), the double-layer had a lower lifecycle than the triple-layer. The failure sequence for both test conditions are predicted based on the failed samples and illustrated in Figure 13.

In the FCT, both coatings are observed to fail at the topcoat/TGO interface, resulting in the topcoat to detach from the bond coat (Figure 7). Owing to the higher thermal coefficient of expansion (CTE) of the metallic substrate, the metallic substrate expands or contracts more than the topcoat. During the heating cycle, the columnar gaps existed in the topcoat will start widening as a form of stress relieve in the topcoat. Due to the low fracture toughness of the GZ (Table 2), lateral cracks were initiated mainly in the SPS GZ layer. For the case of the triple-layer coating, the existing vertical crack in the dense SPPS GZ layer will extend into the underneath layers (i.e., SPS GZ and SPS YSZ layers). The surface of the bond coat starts oxidising through the porosities in the topcoat and the columnar gaps or cracks that existed in the coating structure [19]. As a result, a TGO layer starts to grow between the topcoat and the bond coat in both coatings. The undulated TGO indicated that the TGO layer experienced an in-plane compression in the cooling cycle. Due to the thermal mismatch, the expansion and contraction in each cycle leads to the accumulation of strain energy in the topcoat. Consequently, lateral cracks start forming at the topcoat/TGO interface [78].

As soon as the samples are cycled longer in the FCT, the columnar gaps widened further and the lateral cracks in the SPS GZ layer started propagating laterally along the IPBs. The vertical cracks from the dense SPPS GZ layer of the triple-layer coating continued to extend and widen to alleviate the increasing stresses in the coating structure. The bond coat is oxidised further, associated with the thickening effect on the TGO layer. The lateral cracks around the TGO layer continued propagating laterally until the entire topcoat spalled off from the bond coat. By examining the SEM micrographs of the failed samples for both coatings (Figure 8), the topcoat is seen to undergo a degree of sintering while the bond coat is fully depleted in β-phase. At the early stage, the outward diffusion of the alumina leads to the formation of a slow growing α-alumina layer (also known as the TGO layer). This layer acts as a barrier to avoid the outward diffusion of other reactants such as Cr, Co and Ni. As the samples continued in the FCT, the Al composition throughout the bond coat thickness decreased and subsequently leads to β-phase depletion. The oxygen activity at the TGO interface increases owing to the decrement in Al activity, creating a gradient of oxygen across the TGO layer and encouraging the outward diffusions of other reactants into the TGO layer [49]. Due to the high oxygen activity at the top surface of the TGO layer, spinel is seen to form mostly on top of the TGO layer. As a consequence, more stresses are induced into the topcoat, causing the TBC to fail [49]. Furthermore, Tang et al. suggested that the oxidation of the surface of bond coat may have occurred during the thermal spray process of the HVOF bond coat, which can lead to an early formation of spinel and other oxides [79]; however, the oxidation of bond coat is not investigated in this study.

For the case of BRT, the samples failed at the TGO/topcoat and the SPS GZ/SPS YSZ interfaces. Both coatings experienced a shorter and faster heating and cooling cycle (5 mins heating and 2 mins cooling) as compared to the FCT test. Although the exposure time is much lower than the FCT, the samples are subjected to a much higher surface temperature. Interestingly, the thickness of the TGO layer, measured to be $1.5 \pm 0.5$ μm, is significantly lower than the samples tested in the FCT. The relatively shorter heating time (5 mins) suppresses the oxidation of the bond coat, thereby restricting the growth of the TGO layer. A similar finding is also reported by Mahade et al. [19,25].

In Figure 9(c) and 9(d), it can be observed that the GZ layers (either SPS GZ or SPPS GZ layers) delaminated layer-by-layer, revealing the SPS YSZ layer. The possible reason for the occurrence is mainly due to the low fracture toughness of the GZ (i.e., 0.46 MPa·m$^{-1/2}$ for SPS GZ and 0.58 MPa·m$^{-1/2}$ for SPPS GZ), evidenced by the interlinking horizontal and vertical cracks. As a result, the propagation of lateral cracks in the GZ layer happens more easily as compared to the SPS YSZ layer. After the initiation of the lateral crack, the presence of the IPBs in the coating structure will speed up the propagation by providing a less resistant path to the adjacent cracks or columnar gaps. As the number of cycle increases, these interlinked cracks induced delamination, causing the SPS GZ or the SPPS GZ layer to delaminate layer-by-layer until it revealed the underneath layer.

The higher microhardness value observed in the BRT exposed samples (Table 3) indicated that the topcoat stiffened up at elevated temperatures. The significant increase in microhardness values for samples in BRT compared to FCT was expected due to the more severe sintering effect resulting from the harsher testing environment in BRT; However, it is not clearly understood if sintering was the main cause for the failure of both types of TBCs in this study as SEM micrographs were taken after failure occurred. Interestingly, Mahade et al. also reported that the topcoat has undergone to a certain extent of sintering at the top surface when subjected to a similar BRT test but at a lower surface temperature (1300 °C) [21].

After the spallation of the GZ layer, lateral cracks now appear at two different locations, (a) at the free surface and (b) adjacent to the bond coat. The propagation of the horizontal crack near to the bond coat is slightly more inward to the SPS YSZ layer or at the TGO/topcoat interface, depending on the thermal gradient condition or the interface roughness [80]. The nucleation and propagation of the crack at this region are mainly driven by the CTE mismatch between the topcoat and the substrate. With the formation of TGO, the stresses at the TGO/topcoat interface will be further increased. Subsequently, the propagation of the lateral crack will speed up, leading to a larger part of the coating to falling apart after a period of exposure at high temperatures.

*4.3.CMAS test*

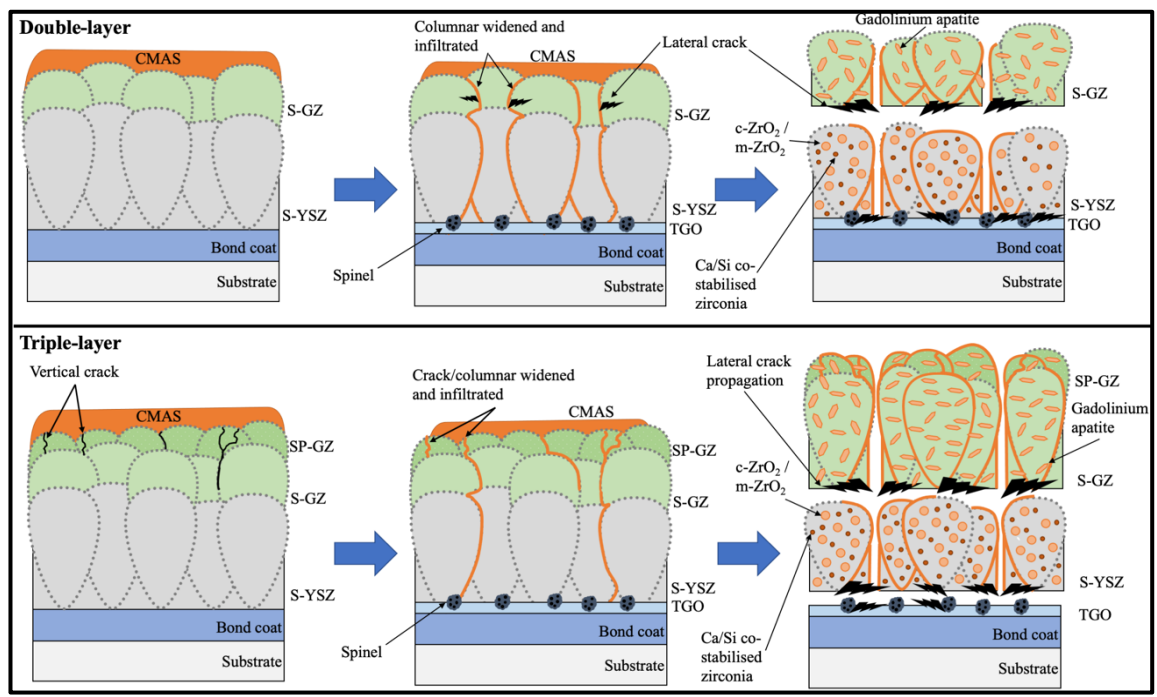

Figure 14. Schematic diagrams of CMAS infiltration in the double-layer and the triple-layer coating. It is worth noticing that the CMAS infiltration is inhibited at the SP-GZ/S-GZ interface, associated to the discontinuation of the columnar feature in the dense GZ layer.

After the CMAS test, both coatings were fully infiltrated with CMAS just after 5 mins of exposure at 1300 °C. The short exposure time is chosen with the idea of investigating how the columnar structured TBC samples will behave under CMAS attack. From the EDX mapping in Figure 11 and Figure 12, it is clearly shown that the columnar gaps act as a pathway for the CMAS to infiltrate the coating. Comparing the double-layer with the triple-layer coating, both coatings were fully infiltrated by CMAS; However, the triple-layer coating is infiltrated via the through channels (4.6 ± 1.5 µm) and partially infiltrated in the minimised intercolumnar gaps or cracks (0.6 ± 0.3 µm) contributed by the dense SPPS GZ layer. It is worth mentioning that the loading condition of the CMAS is chosen according to an industry-standard, tested under an isothermal condition. The predicted failure sequence is illustrated in Figure 14.

Ramping up to the set temperature, potentially 1300 °C, the stress level in the coating structure increases, leading to an increment in the stored elastic strain energy within the coating structure. The columnar gaps in the coating structure will be widened [81]; If the stored strain energy in the coating exceeds the fracture toughness of the coating layer, initiation and propagation of cracks will occur [82]. In the meantime, the pre-deposited CMAS on top of the samples will melt into a molten state and spread all over the top surface of the samples (i.e., double-layer and triple-layer TBCs), forming a liquid reservoir. The molten CMAS starts flowing into the open channels in both the double-layer and triple-layer coatings. The infiltration speed of the CMAS depends on the width of the open channels. A wide-opened channel will ease the flow of CMAS, subsequently infiltrating the entire topcoat within a short amount of time. For the narrow channel, CMAS infiltration speed will be influence by the capillary pull induced by the high capillary pressure as capillary pressure is inversely proportional to the radius of the channel; however, the frictional drag contributed by the contact surfaces along the wall of the channel also plays an important role in slowing down the infiltration speed of CMAS [46,83]. Further investigation will be required to understand how the capillary pressure and frictional drag will affect the CMAS infiltration speed along these narrow channels.

When the molten CMAS comes into contact with GZ, the re-precipitation of GZ will happen almost instantaneously to form gadolinium apatite phase ($Gd_8Ca_2(SiO_4)_6O_2$) [43]. Due to the columnar gaps being too wide in the double-layer coating, the molten CMAS is expected to infiltrate deeper into the coating until it completely reacts. However, in the case of the triple-layer coating, CMAS residues can be found on the surface of the coating and at the SPPS GZ/SPS GZ interface (Figure 12), particularly at the top surface of partially infiltrated channels, suggesting that the CMAS reacted with GZ to form the apatite phase, thereby inhibiting deeper infiltration in these regions. The finding is found to be similar to the study by Kumar et al. suggested that sealing can only occur if the open cavity is less than 1 µm. [29]. Additionally, Krämer et al. [30] reported that the molten

CMAS can take less than 1 min to fully infiltrate an EB-PVD processed YSZ TBC at a temperature just above the melting point of the CMAS used, 1240 °C. Wellman et al. also suggested that a minimum of 4.8 mg/cm$^2$ would be sufficient to cause an EB-PVD TBC to lose the ability to protect the underlying substrate [37]. When the CMAS reaches the SPS YSZ layer, the re-precipitation of the zirconia metastable tetragonal phase will begin, forming yttria-depleted zirconia grains. Additionally, some of these Y-depleted zirconia grains will react with the residual glassy CMAS left in the coating, leading to the repreciptiation of Ca/Si co-stabilised zirconia [30]. Upon cooling, these yttria-depleted zirconia grains will be transformed into the monoclinic (m) zirconia phase which is detrimental to the lifetime of TBCs [34,41].

Due to the high hardness and low CTE of CMAS (6.14 ± 0.1 GPa [84]; 9.32 x 10$^{-6}$ K$^{-1}$ [84], respectively), the CMAS infiltrated coating structure resists any shrinkage in the cooling stage, causing an increase in mechanical stresses that leads to the initiation and propagation of cracks [70]. The increase in these stresses is reflected in the microhardness value measured for CMAS exposed samples, which are found to be significantly increased (Table 3). Besides, the CTE mismatch between the CMAS residue and the topcoat can further contribute to the formation of cracks within the coating structure, as depicted in Figure 11 and Figure 12. When the samples are cooled to room temperature, the CTE mismatch between the topcoat and substrate will also lead to a further increase in the stress level of the coating. As a consequence, the stress level in the coating structure is expected to be way above the fracture toughness in each layer, inducing and propagating cracks along the voids and microcracks in the coating structure until delamination occurs.

## 5. Conclusions

Both the double-layer and the triple-layer coatings were successfully produced through the axial plasma spray, utilizing the suspension and the solution precursor feedstocks. Suspension coating structures represent the typical columnar structure that offers excellent strain tolerance capability in thermal cycling. The dense SPPS GZ layer represents an intermediate structure, between the vertical crack and columnar structure and it is well-attached on top of the SPS GZ layer with no apparent pores or cracks at the interface. Most of the intercolumnar gaps in the SPPS GZ layer are minimised. The success in minimising the intercolumnar gaps increases the CMAS resistance without compromising the strain-capability of the coating structure. The thermal cycling lifetime and the CMAS attack for both the coatings were compared. The following conclusions can be drawn:

- In the FCT, both the double- and the triple-layer coatings show a similar thermal cycling lifetime; In the BRT, the triple-layer coating performed better than the double-layer. The failure mode in the FCT test differs from the failure mode in the BRT.

- In the FCT test, the failure mechanisms of the topcoat is dominated by the CTE mismatch between the topcoat and the substrate and the thickening effect of the TGO layer. Failure occurred at the topcoat/TGO interface.

- In the BRT test, the samples were subjected to a sudden thermal load due to the short exposure time at high surface temperatures of 1360 °C. The interlinking of horizontal and vertical cracks in the GZ layer is mainly due to the lower fracture toughness of GZ m In contrast, the lateral cracks propagated along the IPBs, which is believed that the IPBs acted as a pathway to help the propagation of the lateral cracks. The disappearance of the GZ layer suggested that the GZ layer delaminated layer-by-layer until it revealed the SPS YSZ layer. The latter failure mode is observed at the topcoat/TGO interface, where delamination occurred due to the CTE mismatch between the topcoat and the substrate.

- Both the coatings were fully infiltrated with CMAS. In the double-layer, CMAS is mainly detected around the columnar gaps. It is believed that the columnar gaps are too wide for the GZ to react with CMAS to form a sealant layer; whereas for the case of the triple-layer coating, the topcoat is fully infiltrated along the through channels (developed from the SPS YSZ layer to the top surface of the topcoat) and partially infiltrated along the minimised columnar features or cracks and stopped at the SPPS GZ/SPS GZ interface. The infiltration depth of these partial infiltrated columns or cracks is within the range of 38 μm to 68 μm. The width of the infiltrated through channels is measured to be $4.6 \pm 1.5$ μm, while the width of the partial infiltrated channel is measured to be $0.6 \pm 0.3$ μm.

## Acknowledgements

This work was supported by the Engineering and Physical Sciences Research Council (ESPRC) (grant number EP/R511730/1). The authors would like to thank John Kirk for his assistance during the HVOF thermal spray. The authors acknowledge the Nanoscale and Microscale Research Centre (nmRC) at the University of Nottingham for providing access to SEM and FEG-SEM facilities (grant number EP/L022494/1).